\pdfoutput=1
\documentclass[reprint, amsmath,amssymb, aps, prd, longbibliography, superscriptaddress]{revtex4-1}

\usepackage{listings}
\usepackage{booktabs}
\usepackage{graphicx}
\usepackage{amsmath}
\usepackage{braket}
\usepackage[frozencache]{minted}
\usepackage{caption}
\usepackage{subcaption}
\usepackage{multirow}
\usepackage[colorlinks=true,urlcolor=blue,citecolor=blue,linkcolor=blue]{hyperref}
\usepackage{xcolor}
\usepackage[htt]{hyphenat}  

\newcommand{\qmap}{\mathcal{E}}
\DeclareMathOperator{\Tr}{Tr}

\DeclareMathOperator{\opy}{\hat{\sigma_y}}

\newcommand{\ketbra}[2]{\mbox{$\left|#1\right\rangle\!\!\left\langle #2\right|$}}

\begin{document}

\title{Qiskit Pulse: Programming Quantum Computers Through the Cloud with Pulses}
\author{Thomas Alexander}
\thanks{These two authors contributed equally}
\email[\\Corresponding author: ]{talexander@ibm.com}
\affiliation{IBM T.J.~Watson Research Center, Yorktown Heights, NY 10598, USA}

\author{Naoki Kanazawa}
\thanks{These two authors contributed equally}
\email[\\Corresponding author: ]{talexander@ibm.com}
\affiliation{IBM Research - Tokyo, 19-21 Nihonbashi Hakozaki-cho, Chuo-ku, Tokyo, 103-8510, Japan}

\author{Daniel J. Egger}
\affiliation{IBM Research - Z\"urich, S\"aumerstrasse 4, 8803 R\"uschlikon, Switzerland}

\author{Lauren Capelluto}
\affiliation{IBM T.J.~Watson Research Center, Yorktown Heights, NY 10598, USA}

\author{Christopher J. Wood}
\affiliation{IBM T.J.~Watson Research Center, Yorktown Heights, NY 10598, USA}

\author{Ali Javadi-Abhari}
\affiliation{IBM T.J.~Watson Research Center, Yorktown Heights, NY 10598, USA}

\author{David McKay}
\affiliation{IBM T.J.~Watson Research Center, Yorktown Heights, NY 10598, USA}



\begin{abstract}

The quantum circuit model is an abstraction that hides the underlying physical implementation of gates and measurements on a quantum computer. For precise control of real quantum hardware, the ability to execute pulse and readout-level instructions is required. To that end, we introduce Qiskit Pulse, a pulse-level programming paradigm implemented as a module within Qiskit-Terra \cite{Qiskit}.
To demonstrate the capabilities of Qiskit Pulse, we calibrate both un-echoed and echoed variants of the cross-resonance entangling gate with a pair of qubits on an IBM Quantum system accessible through the cloud.
We perform Hamiltonian characterization of both single and two-pulse variants of the cross-resonance entangling gate with varying amplitudes on a cloud-based IBM Quantum system. We then transform these calibrated sequences into a high-fidelity CNOT gate by applying pre and post local-rotations to the qubits, achieving average gate fidelities of $F=0.981$ and $F=0.979$ for the un-echoed and echoed respectively. This is comparable to the standard backend CNOT fidelity of $F_{CX}=0.984$.
Furthermore, to illustrate how users can access their results at different levels of the readout chain, we build a custom discriminator to investigate qubit readout correlations.
Qiskit Pulse allows users to explore advanced control schemes such as optimal control theory, dynamical decoupling, and error mitigation that are not available within the circuit model.

\end{abstract}

\maketitle

\section{Introduction}

In quantum computing, information is stored and processed according to the laws of quantum mechanics~\cite{Nielsen2000}.
The primary quantum programming paradigm is the circuit model.
In this model, the underlying dynamics of the physical system implementing the quantum computer are abstracted as a sequence of unitary gate operations and projective measurements applied to a set of qubits.
Gates manipulate the states of qubits, while measurements extract classical information in the form of bit-strings, which encode the outcome of projective measurements of the qubits in a particular measurement basis.

Qiskit is an open-source quantum computing framework designed to enable research on near-term quantum computers and their applications. It provides tools for creating, manipulating and running quantum programs on quantum systems independent of their underlying technology and architecture. The standard programming abstraction for a quantum circuit is a quantum assembly language (QASM) such as OpenQASM \cite{Cross2017} which Qiskit supports \cite{Qiskit} and many similar languages that have been described in the literature \cite{chuangQasm2circ, khammassi2018}. However, hardware is not capable of natively implementing quantum instructions and must compose these operations from the classical stimulus avilable to control hardware.

At the hardware level the time-dependent dynamics of a quantum system interacting with applied control fields is described by its Hamiltonian and the  Schr\"{o}dinger equation.
Through careful engineering of applied classical control fields a quantum system may be steered through a desired unitary evolution \cite{Glaser2015}.
Superconducting transmon qubits, for example, encode a qubit in a non-linear oscillator formed by a parallel circuit consisting of a Josephson junction and capacitor, and may be manipulated by applying shaped microwave control pulses~\cite{Koch2007}.
Implementing the quantum circuit model on such an architecture requires compiling circuit instructions to a set of microwave control instructions, or pulses, which enact the desired state-transformations and/or measurements.

In the circuit domain, an atomic circuit instruction is agnostic to its pulse-level implementation on hardware. Extracting the highest performance out of quantum hardware requires the ability to craft a pulse-level instruction schedule, which cannot be done within the standard circuit model.
To enable pulse-level programming an instruction set, OpenPulse \cite{Mckay2018}, was developed to describe quantum programs as a sequence of
pulses, scheduled in time. We present within this paper a Python implementation of OpenPulse, \textit{Qiskit Pulse} which adds to the Qiskit compilation pipeline the capability to schedule a quantum circuit into a pulse program intermediate representation, perform analysis and optimizations, and then compile to OpenPulse object code to execute on a quantum system.

The various hardware architectures used for current-day quantum computing systems creates a need for a pulse-level instruction set that may address most platforms at an abstract level, compatible with both commercial off-the-shelf and proprietary control instruments, including arbitrary waveform generators (AWG), signal generators, filters, amplifiers and digitizers \cite{Krantz2019}.
To program such systems at the pulse-level in a \textit{hardware independent} manner requires the user-level instruction set to be target-compiled to the underlying system hardware components, each of which may have a unique instruction set and programming model.
Recent efforts to construct microarchitectures that conform to classical computer engineering paradigms \cite{Fu2017, Butko2019, Ryan2017} have programming semantics closely tied to the underlying microarchitecture.
Compiling directly from the circuit model to target hardware obfuscates the underlying pulses that manipulate the hardware removing a powerful degree of control.

With Qiskit Pulse we enable the development of a common and reusable suite of technology-independent quantum control techniques \cite{Glaser2015} that operate at the level of analog stimulus which may be remotely retargeted to cloud-based quantum computing systems.

\emph{Paper outline ---} In Sec.~\ref{sec:pulse_model} we present our pulse programming model.
We demonstrate the capabilities of Qiskit Pulse in Sec.~\ref{sec:cr-pulse} where we show Hamiltonian characterization of the two-qubit cross-resonance interaction, and calibration of a high-fidelity entangling gate, on a cloud-based quantum computer available on the IBM Quantum Experience. We discuss how the readout of quantum computers is incorporated in Qiskit Pulse in Sec.~\ref{sec:discriminators} and conclude in Sec.~\ref{sec:conclusion}. The source code and data for the experiments within this paper has been made available online \cite{alexanderNotebooksDataQiskitPulse2020a}.

\section{Qiskit Pulse programming model}
\label{sec:pulse_model}

In the standard quantum circuit model, the time elapsed between operations is irrelevant as long as the order of non-commuting gates is preserved \cite{metodiSchedulingPhysicalOperations2006}.
However, when controlling quantum hardware at the pulse level, properly timing and synchronizing instructions is crucial for accurately enacting quantum operations.
For instance, users may create new gate definitions, characterize and correct for crosstalk on qubits neighboring interacting qubits, implement optimal control techniques such as GRAPE \cite{Khaneja2005} or mitigate errors through Richardson extrapolation \cite{Temme2017, richardsonVIIIDeferredApproach1927, Kandala2019}.

\begin{figure}
  \includegraphics[width=0.9\linewidth]{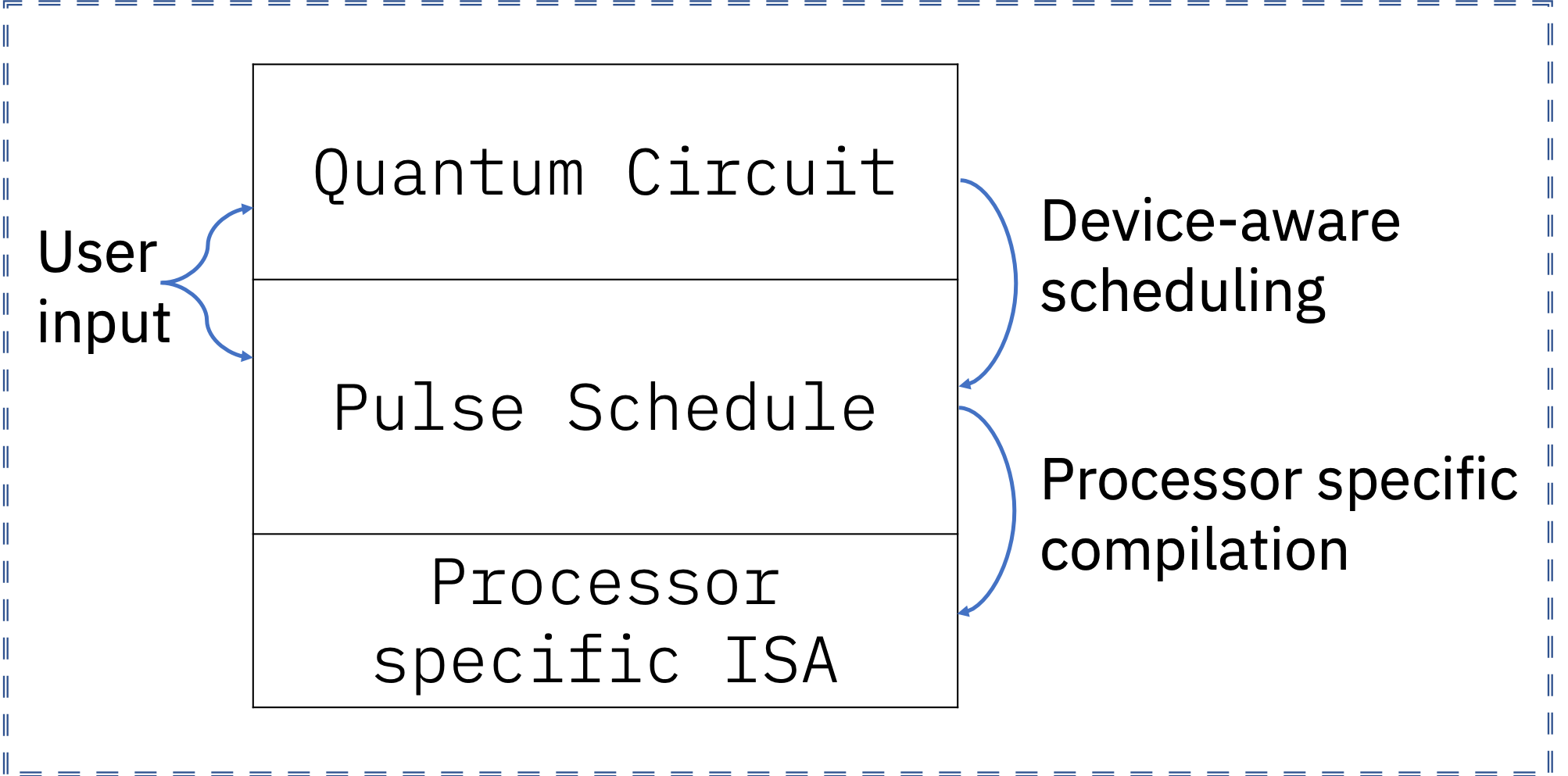}
  \caption{The envisioned quantum program representations and their associated lowering compilation procedures. QASM programs may be built and optimized with information of the system topology, native gates, and error rates, and then are \textit{scheduled} into pulse programs by using calibrated native gate definitions. Pulse programs are compiled to a processor-specific ISA through a target code generation procedure. The typical user is expected to program at the circuit level, whereas Qiskit Pulse enables advanced users to control at the pulse level.}
  \label{fig:programreps}
\end{figure}

We envision that a classical microprocessor with an embedded \textit{pulse coprocessor} will be responsible for controlling and measuring the quantum device. Within this work we only focus on describing a virtual execution model and limited set of instructions for the pulse coprocessor which can be compiled to the instruction set architecture (ISA) of the underlying control hardware. Qiskit Pulse's position in the predicted quantum computing compilation pipeline is demonstrated in Fig.~\ref{fig:programreps}. We expect that as quantum hardware continues to be refined these abstractions will be extended.

Qiskit Pulse provides an open source, front-end implementation of the OpenPulse interface \cite{Mckay2018}. Third parties can fully integrate with Qiskit Pulse by implementing their own Qiskit provider \cite{Qiskit} which is responsible for translating Qiskit Pulse programs to executable programs on the provider-specific hardware which might include components such as AWGs and digitizers.
Qiskit Pulse programs are composed of \emph{pulses}, \emph{channels}, and \emph{instructions} which we present in the following subsections.

\subsection{Pulses}\label{sec:wfs}

A \textit{pulse} is a \textit{time-series} of complex-valued amplitudes with a maximum unit norm, $[d_0, \ldots, d_{n-1}]$. Each $d_j$, $j\in\{0,..., n-1\}$, is called a \textit{sample}.
Every system specifies a cycle-time of \texttt{dt} which is the finest time-resolution exposed on the pulse coprocessor, and is typically defined by the sample rate of the coprocessor's waveform generators.
Each sample in a pulse is output for one cycle, a \textit{timestep}.
All pulse durations and timesteps are defined and discretized dimensionlessly with respect to \texttt{dt}.
The ideal output signal has amplitude
\begin{align} \label{egn:pwcc_pulse}
D_j =  \textrm{Re}\left[e^{i 2\pi f j \texttt{dt} + \phi}d_j \right]
\end{align}
at time $j\texttt{dt}$, where $f$ and $\phi$ are a modulation frequency and a phase.
The pulse samples describe only the envelope of the produced signal which are then mixed up in hardware with a carrier signal defined by a frequency and a phase.
To reduce encoding sizes we also allow hardware providers to define parametric pulse shapes.
For example, one parametric pulse supported by IBM backends and made available through the Pulse library is the \texttt{Gaussian} pulse. It takes three parameters: an integer \texttt{duration} in terms of \texttt{dt}, a complex amplitude \texttt{amp}, and standard deviation \texttt{sigma}. This parametric pulse can be instantiated within Qiskit in the following way:

\begin{minted}[fontsize=\footnotesize, frame=lines,framesep=2mm, autogobble]{python}
from qiskit.pulse.pulse_lib import Gaussian

duration = 128
amp = 0.2
sigma = 16
gaussian_pulse = Gaussian(duration, amp, sigma)
\end{minted}

\begin{table*}
\centering
\caption{\label{tab:instructions} A summary of channels and the pulse instructions that are defined on them.
Note that the \texttt{DriveChannel}, \texttt{MeasureChannel}, and \texttt{ControlChannel} are subtypes of \texttt{PulseChannel}, whereas the \texttt{AcquireChannel}, which cannot transmit stimulus pulses, is not. In the near term it is expected that the instruction set below will continue to be expanded.}
\begin{tabular}{c}

\begin{tabular}{ p{28mm} p{44mm} p{88mm} c}\hline\hline
Channel	& Alias & Description\\\hline
\texttt{PulseChannel} & - & Generic transmit channel used to manipulate the quantum system. Current supported sub-types are \texttt{DriveChannel}, \texttt{MeasureChannel} and \texttt{ControlChannel}. \\\hline
\texttt{DriveChannel} & $d_i$ &  Transmit channel connected to qubit $i$, with signals typically modulated at a frequency in resonance with qubit $i$.\\\hline
\texttt{MeasureChannel} & $m_i$ & Transmit channel connected to the readout component of qubit $i$.\\\hline
\texttt{ControlChannel} & $u_i$ & Transmit channel with signals typically associated with arbitrary interaction terms in the Hamiltonian.\\\hline
\texttt{AcquireChannel} & $a_i$ & Receive channel connected to the readout component of qubit $i$, capable of digitizing and acquiring data.\\
\end{tabular}

\\

\begin{tabular}{ p{28mm} p{44mm} p{88mm} c}\hline\hline
Instruction & Operands & Description\\\hline
\texttt{Play} & \texttt{pulse:} \texttt{Pulse}, \newline \texttt{channel:} \texttt{PulseChannel} & Output the waveform described by \texttt{pulse} on the \texttt{channel}.\\\hline
\texttt{Delay} & \texttt{duration:} \texttt{int}, \newline \texttt{channel:} \texttt{Channel} & Idle the \texttt{channel} for the given \texttt{duration}. \\\hline
\texttt{ShiftPhase} & \texttt{phase:} \texttt{float},\newline \texttt{channel:} \texttt{PulseChannel} & Shift the phase of the \texttt{channel} by \texttt{phase} radians. \\\hline
\texttt{SetFrequency} & \texttt{frequency:} \texttt{float},\newline \texttt{channel:} \texttt{PulseChannel} & Set the frequency of the \texttt{channel} to \texttt{frequency} Hz. \\\hline
\texttt{Acquire} & \texttt{duration:} \texttt{int},\newline \texttt{channel:} \texttt{AcquireChannel},\newline \texttt{register:} \texttt{Register}	&  Trigger the \texttt{channel} to collect data for the given \texttt{duration}, and store the measurement result in a \texttt{register}. \\\hline\hline
\end{tabular}

\end{tabular}
\end{table*}

\subsection{Channels}
Hardware components are modeled with \texttt{Channel}s. Channels label signal lines that either transmit or receive signals between the control electronics and the quantum device. Each channel executes instructions from a first-in, first-out (FIFO) queue as outlined in subsection \ref{sec:instructions}.

Channels are constrained at target code generation time to target hardware components, e.g., an AWG.
The calibrated parameters of a channel, such as its frequency, and the pulses played on that channel, depend on the physical properties of the targeted qubit.
Therefore, channels are not interchangeable at the pulse abstraction layer, i.e. permuting channels over qubits will not give equivalent results.
This highlights another difference between circuit and pulse instructions, the parameters of gates in a quantum circuit do not depend on the physical properties of the targeted qubits.
The qubits in a quantum circuit can therefore be interchanged without affecting the computational result as long as the topology of the device is taken into account and gate imperfections are ignored.

There are several different channel types, and each may support a different instruction set. A summary of channels and their descriptions is provided in Table.~\ref{tab:instructions}. The channel type, and thus the supported instructions, is determined by the effect of the channel on the quantum device. For example, a \texttt{PulseChannel} models the output of a control field $\alpha_k(t)$ in a system Hamiltonian $\hat H(t)=\hat H_\text{sys}+\sum_k \alpha_k(t)\hat H_k$ which is composed of time-independent system and time-dependent control terms. The Hamiltonian term for a given channel, $\hat H_k$ in general may be arbitrary, but is typically associated with the subtype of the respective pulse channel that is assigned at system configuration time.

The sub-types of \texttt{PulseChannel}s are \texttt{DriveChannel}s, \texttt{MeasureChannel}s, and \texttt{ControlChannel}s.
Each pulse channel maintains an instruction writeable frequency $f$ and phase $\phi$, which modify the channel output as per Eq.~(\ref{egn:pwcc_pulse}).
Tracking the phase in this way enables the implementation of \textit{virtual} Z-gates \cite{lurieNumericalDesignComposite1986, Mckay2017}.
Qubit drive and readout pulses are respectively assigned to \texttt{DriveChannel}s and \texttt{MeasureChannel}s, see e.g., the drive pulses on drive channels \texttt{d0} and \texttt{d1} in Fig.~\ref{fig:circtosched}.
Their index is trivially mapped to the address of the target qubit.
The \texttt{ControlChannel} implements any remaining control fields, such as coupler drives or two-qubit drives as depicted by \texttt{u2} in Fig.~\ref{fig:circtosched}.
The backend hardware may choose to map multiple \texttt{PulseChannel}s to the same control unit in the system, which enables tracking a unique phase for each channel.
For example, in the IBM Quantum systems used to perform the experiments within this paper, every \texttt{DriveChannel} may share an AWG with multiple \texttt{ControlChannel}s, each of which have a frequency and phase adjusted to track that of their respectively coupled qubits, enabling the implementation of two-qubit gates as demonstrated in Sec. \ref{sec:cr-pulse}.

The \texttt{AcquireChannel} is used to communicate to the system when qubit readout data must be acquired.
It is not associated with a control term in the Hamiltonian and does not output stimulus to the quantum system.
Data collected on these channels are used to determine the qubit state, see Sec. \ref{sec:discriminators} for more details.

For convenience we alias drive, measurement, control and acquisition channels as $\{d_0, \ldots, d_{n_{q}-1}\}$, $\{m_0, \ldots, m_{n_{q}-1}\}$, $\{u_0, \ldots, u_{n_{u}-1}\}$, and $\{a_0, \ldots, a_{n_{q}-1}\}$ respectively, where $n_q$ is the number of qubits and $n_u$ is the number of arbitrary control channels of the system.

\subsection{Instructions and Execution Model}
\label{sec:instructions}
\texttt{Instruction}s may be scheduled on \texttt{Channel}s to manipulate the quantum system.
Pulse instructions have as operands channels and instruction-dependent constants.
All pulse instructions have a fixed, deterministic duration, which may be specified either implicitly or explicitly.
Instructions are executed with an \textit{allocation} and \textit{trigger} timing model in which instructions are loaded into a FIFO queue unique to each channel and then execution is initialized synchronously across all channels with an external trigger signal.
Consequently the absolute start and end of every pulse instruction may be scheduled at compile-time across channels with hard real-time deadlines relative to the external trigger signal.
Instructions may have multiple channels as operands causing an execution dependency. In this case channels stall execution until all operand channels are available.
We now outline the different types of instructions, which are summarized in Table \ref{tab:instructions}.

Every channel supports a \texttt{Delay} instruction which has as operands a \texttt{duration} which is specified as a number of cycles, and a target channel. The channel will idle for the duration of the instruction.

The \texttt{Play} instruction allows users to play a \texttt{pulse} on a target \texttt{PulseChannel} with a frequency and phase set with the \texttt{ShiftPhase} and \texttt{SetFrequency} instructions.
The \texttt{ShiftPhase} instruction has an implicit duration of zero and accepts an input float \texttt{phase} and \texttt{PulseChannel}.
The \texttt{ShiftPhase} will shift the phase $\phi$ of the target channel by \texttt{phase} radians.
This relative shift persists on the channel from the time of the instruction, allowing the phase of a channel to be accumulated throughout an experiment. The \texttt{SetFrequency} instruction has an implicit duration of zero and accepts an input float \texttt{frequency} and a \texttt{PulseChannel}.
This instruction will set the frequency $f$ of all proceeding pulses on the target channel to \texttt{frequency} Hz.
Like any other instruction, \texttt{SetFrequency} can be used on a single channel multiple times within a schedule, subject to the instantaneous bandwidth of the hardware.
It can therefore be used, for example, to measure the anharmonicity of a transmon qubit.

The \texttt{Acquire} instruction has as operands a \texttt{duration}, an \texttt{AcquireChannel}, and a classical \texttt{register} in which to store the observed result.
This instruction signals to the measurement unit to begin acquiring data, and for how long. Each \texttt{Acquire} instruction should align with a corresponding measurement stimulus \texttt{Play} instruction to induce a measurement of the target qubit.
An acquisition channel outputs an unsigned integer value \texttt{N} into the result register. For the standard two-level qubit, this will be a single bit $\{0,1\}^1$.

If a measurement stimulus pulse measures multiple qubits, as is typical for multiplexed measurement schemes~\cite{Jeffrey2014, Heinsoo2018}, an acquisition instruction must be synchronously scheduled for each of the measured qubits.
In hardware, each \texttt{AcquireChannel} is constrained to a measurement chain which usually includes data acquisition, filtering, kerneling, and state discrimination.
To accommodate the heterogeneous readout schemes encountered in hardware we defined three levels of readout data and how to convert between them, see Sec.~\ref{sec:discriminators}.

The set of operations in Qiskit Pulse should have sufficient generality to program a pulse coprocessor for an arbitrary quantum computing system in the time-domain and be embedded within a larger instruction set that might include both classical control flow and a traditional gate-level description of a quantum program with instructions being implemented by a lowering procedure to pulse instructions. The benefit of this approach is that a single software stack may provide the middle-end for the rapidly developing heterogeneous quantum computing platforms.

\subsection{The Pulse Schedule}
The pulse \texttt{Schedule} is the representation of a pulse program in Qiskit Pulse and is an ordered collection of scheduled pulse instructions. The pulse schedule is equivalent to a basic block \cite{cooperEngineeringCompiler2003} in a classical computation with deterministic instruction durations. To construct a pulse schedule \texttt{Instruction}s may be appended as demonstrated in the example below, which prepares qubit 0 in the $\ket{1}$ state and then measures it:

\begin{minted}[fontsize=\footnotesize, frame=lines,framesep=2mm, autogobble]{python}
  # Create a pulse schedule.
  sched = Schedule(name='excited_state')

  # Create gate and measurement pulses.
  x180 = Drag(x_dur, x_amp, x_sigma, x_beta)
  measure = GaussianSquare(m_dur, m_amp, m_sigma, m_square_width)

  # += appends an Instruction to a Schedule.
  sched += Play(x180, DriveChannel(0))

  # Measure qubit 0.
  sched += Play(measure, MeasureChannel(0))

  # Determine the state of qubit 0 and store it
  # in a persistent MemorySlot register which
  # will be returned in the program result.
  sched += Acquire(AcquireChannel(0), MemorySlot(0))

  # Run the schedule and get the result.
  counts = execute(sched, backend).result().get_counts()

\end{minted}

\subsection{Scheduling}

Quantum circuits and pulse schedules are both representations of a quantum program.
The Qiskit \textit{transpiler} optimizes quantum circuits according to the properties of the targeted quantum system such as the device topology, the native gate set, and the gate fidelities.
A \textit{scheduler} compiles a circuit program to a pulse program, as depicted in Fig.~\ref{fig:programreps}.
Scheduling requires system dependent information, most notably the definitions of the native gates in terms of scheduled pulse instructions. The scheduler therefore requires a quantum circuit to be transpiled to the native gate set of the target system prior to scheduling.
Furthermore, during scheduling it is crucial to maintain the relative timing of groups of pulse instructions calibrated to implement an element of the native gate set.
For instance, the cross-talk cancellation tone of a cross-resonance gate applied to the target qubit must be aligned with the pulse that drives the control qubit at the frequency of the target qubit \cite{Sheldon2016b}.

\begin{figure}[htbp!]
  \begin{subfigure}[t]{0.45\textwidth}
    \caption{}
    \begin{minted}[fontsize=\footnotesize, frame=lines,framesep=2mm, autogobble]{python}
        qc = QuantumCircuit(2, 2)
        qc.h(1)
        qc.cx(1, 0)
        qc.measure([0, 1], [0, 1])
        qc = transpile(qc, backend)
        pulse_schedule = schedule(qc, backend)

        # Plot the program representations.
        qc.draw()
        pulse_schedule.draw()
    \end{minted}
  \end{subfigure}

  \begin{subfigure}[t]{1.0\linewidth}
    \includegraphics[width=0.95\linewidth]{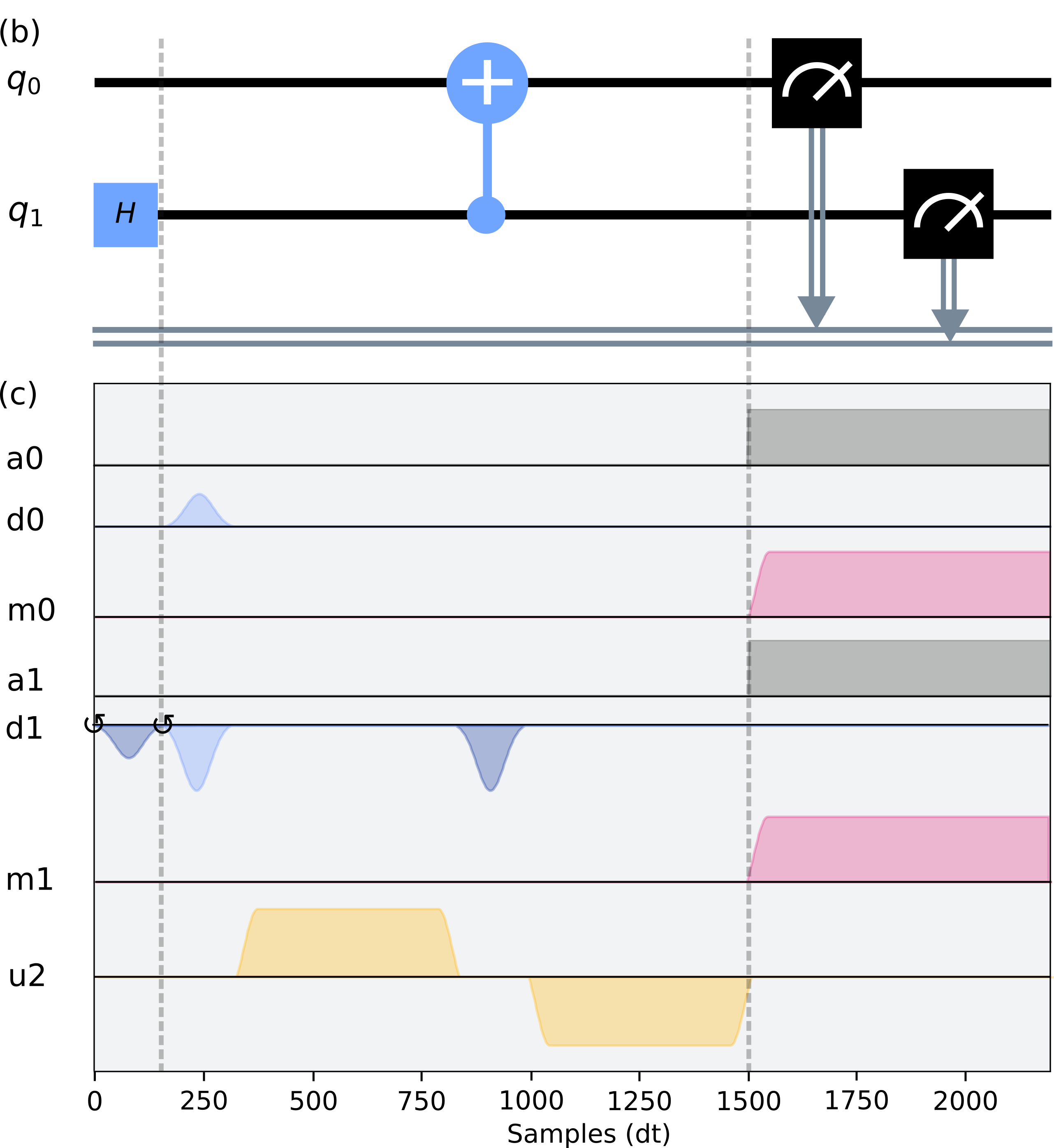}
  \end{subfigure}
  \caption{(a) Qiskit code to construct a quantum circuit that prepares and measures a Bell state and then schedules the circuit to produce an equivalent pulse schedule. Here, \texttt{h} is a Hadamard gate, \texttt{cx} a CNOT gate and \texttt{backend} is a description of a quantum system received from a hardware provider. (b) and (c) Visualization of the mapping between circuit instructions (b) and the composite pulse sequences that will implement the circuit elements (c).
  Pulse envelopes filled with bright and dark colors respectively represent the real (in-phase) and imaginary (quadrature-phase) components of the input control waveform. The circular arrows represent a phase shift.
  The gray shadow on \texttt{a0} and \texttt{a1} indicates the data acquisition trigger for the ADC which is synchronized with the measurement stimulus pulse.
  These mappings are automatically provided by the hardware backend, but may be overridden by the user as we demonstrate in section \ref{sec:cr-pulse}. The scheduler will align the gates in time according to the selected scheduling policy.}
  \label{fig:circtosched}
\end{figure}

The input circuit provides only implicit topological timing constraints, which allows the scheduler to arbitrarily resolve the remaining free time-alignment parameters in the output schedule.
The scheduler's behavior in resolving free parameters is set by specifying a \textit{scheduling method} or \textit{policy}.
By default the Qiskit scheduler follows an ``as-late-as-possible" scheduling method \cite{Qiskit}.
This will schedule individual gates as late as possible while minimizing the deadtime between instructions on the same channel.
This scheduling routine mitigates $T_1$ and $T_2$-decay errors by maximizing the time that qubits will spend in their initial ground state prior to the first pulse, while also minimizing the time between the last pulse and the measurement.
Fig. \ref{fig:circtosched} provides a code snippet for scheduling a quantum circuit into a pulse schedule using Qiskit, and visually demonstrates the correspondence between the circuit instructions input to the scheduler and the calibrated output pulse sequences. This output is easily generated for both the \texttt{QuantumCircuit} and \texttt{Schedule} with the \texttt{draw} method.

Qiskit Pulse users may create pulse programs to replace the default pulse programs of the native gate set provided by the backend and pass them as an argument to the scheduler. This gives users low level control over the gate definitions used at scheduling time. They may specify their own scheduling policies to dynamically aggregate gates and generate composite pulse sequences such as would be required to implement the compilation techniques described by Shi et. al. \cite{shiOptimizedCompilationAggregated2019a}.

\section{Demonstration of a Cross-Resonance Entangling Gate}
\label{sec:cr-pulse}


To highlight how Qiskit Pulse can enable tasks that cannot be done in the circuit model we perform standard quantum process tomography (QPT)~\cite{Mohseni2008} of both echoed and un-echoed cross-resonance (CR)~\cite{Chow2011} pulses for \textit{varying amplitudes} on a cloud-based quantum computer. We use the tomography data to calculate the coefficients of the effective CR Hamiltonian as a function of the pulse amplitude, and show how to implement a high-fidelity Controlled-NOT (CNOT) gate based on the calibrated CR pulse.

\subsection{The Cross-Resonance Interaction}

The CR gate is a microwave-only two-qubit entangling gate for fixed-frequency dispersively coupled qubits~\cite{Chow2011}. It is physically realized by driving the control qubit with microwave pulses at the frequency of the target qubit to stimulate the evolution of an effective $ZX$ interaction Hamiltonian, where $Z$ and $X$ are the Pauli-$Z$ and $X$ operators of the driven control qubit and the target qubit, respectively.

The two-transmon system driven by the CR pulse is described by a time-dependent Hamiltonian $H_{\text{CR}}(t)$ which, in the absence of noise, results in the unitary evolution $U_{\text{CR}}$.
We can approximate the evolution as being generated by a time-independent Hamiltonian $\overline{H}_{\text{CR}}$ using the perturbative method presented in Ref~\cite{Magesan2018} which showed good agreement with experimental results for single-pulse and echoed CR gates \cite{Sheldon2016b}.
This technique approximates the time-dependent control qubit drive pulse with a constant amplitude pulse and block-diagonalizes the resulting Hamiltonian to second order.
Using this approach the CR evolution is approximated by $U_{\text{CR}} \approx \exp\left( -i t_{\text{CR}}\overline{H}_{\text{CR}}\right)$ where
\begin{align}
 \overline{H}_{\text{CR}} &= \frac{Z\otimes B}{2} + \frac{I \otimes C}{2}\label{eqn:hcr} \\ \notag
 B &= \omega_{ZI} I + \omega_{ZX} X + \omega_{ZY} Y + \omega_{ZZ} Z \\ \notag
 C &= \omega_{IX} X + \omega_{IY} Y + \omega_{IZ} Z.
 \end{align}
If the $ZX$ term could be isolated, the resulting unitary gate would be a two-qubit rotation
\begin{equation}\label{eqn:ideal_cr}
U_{\text{ZX}}(\theta_{ZX}) = \exp\left(-i \theta_{ZX} \frac{ZX}{2}\right),
\end{equation}
where the rotation angle $\theta_{ZX}$ depends on the strength and duration of the pulse applied to the control qubit. The unitary gate $U_{\text{ZX}}(\pi / 2)$ is a \emph{perfect entangler} -- it can be used to generate a maximally entangled state from a separable input state and is locally equivalent to a CNOT gate~\cite{Zhang2003}.
Therefore, combined with arbitrary single-qubit operations, it is sufficient for universal quantum computation.

The terms in addition to $\omega_{ZX} ZX$ in $\overline{H}_{\text{CR}}$ lead to coherent errors and divergences from the ideal target unitary in Eq.~\eqref{eqn:ideal_cr}. Characterizing the strength of these terms and designing pulse sequences that suppress them is necessary to create high fidelity entangling operations.
The standard techniques used to suppress these terms are multi-pulse echos and cancellation tones \cite{Sheldon2016b}.

\subsection{Constructing and Calibrating a Cross-Resonance Gate}\label{sec:calib-cr-pulse}

The experiments presented within this section are executed on the twenty-qubit IBM Quantum system \texttt{ibmq\_almaden} to take advantage of higher resolution waveforms with a cycle-time $\texttt{dt}=0.222~\rm{ns}$ afforded by infrastructure under test on that system at the time of writing.
We use \mbox{qubit 1} and \mbox{qubit 0} as the control and target qubits, respectively.
The resonance frequency and anharmonicity of the control qubit are $f_1 = 4.972~\rm{GHz}$ and $\delta_1 = -319.7~\rm{MHz}$, and $f_0 = 4.857~\rm{GHz}$ and $\delta_0=-320.2~\rm{MHz}$ for the target qubit.

We implement both a single-pulse (CR1), and an echoed two-pulse (CR2) variant of the CR gate without a cross-talk cancellation tone on the target qubit \cite{Willsch2017}.
The CR pulse envelope is a \texttt{GaussianSquare} pulse, i.e. a square pulse with Gaussian-shaped rising and falling edges.
The pulse has a square amplitude $A$, a phase $\phi = -0.166~\rm{rad.}$, discussed in Appendix \ref{sec:phase_cal}, and a total duration $t_\text{CR} = 848~\texttt{dt} = 184.4~\rm{ns}$.
The square portion of the pulse has a duration of 720~\texttt{dt} and the Gaussian rising and falling edges last 64~\texttt{dt} and have a 32~\texttt{dt} standard deviation.
The pulse duration is chosen so that a $\pi/2$ rotation angle can be achieved within the weak driving regime.

\begin{figure}[htbp!]
  \includegraphics[width=0.95\linewidth]{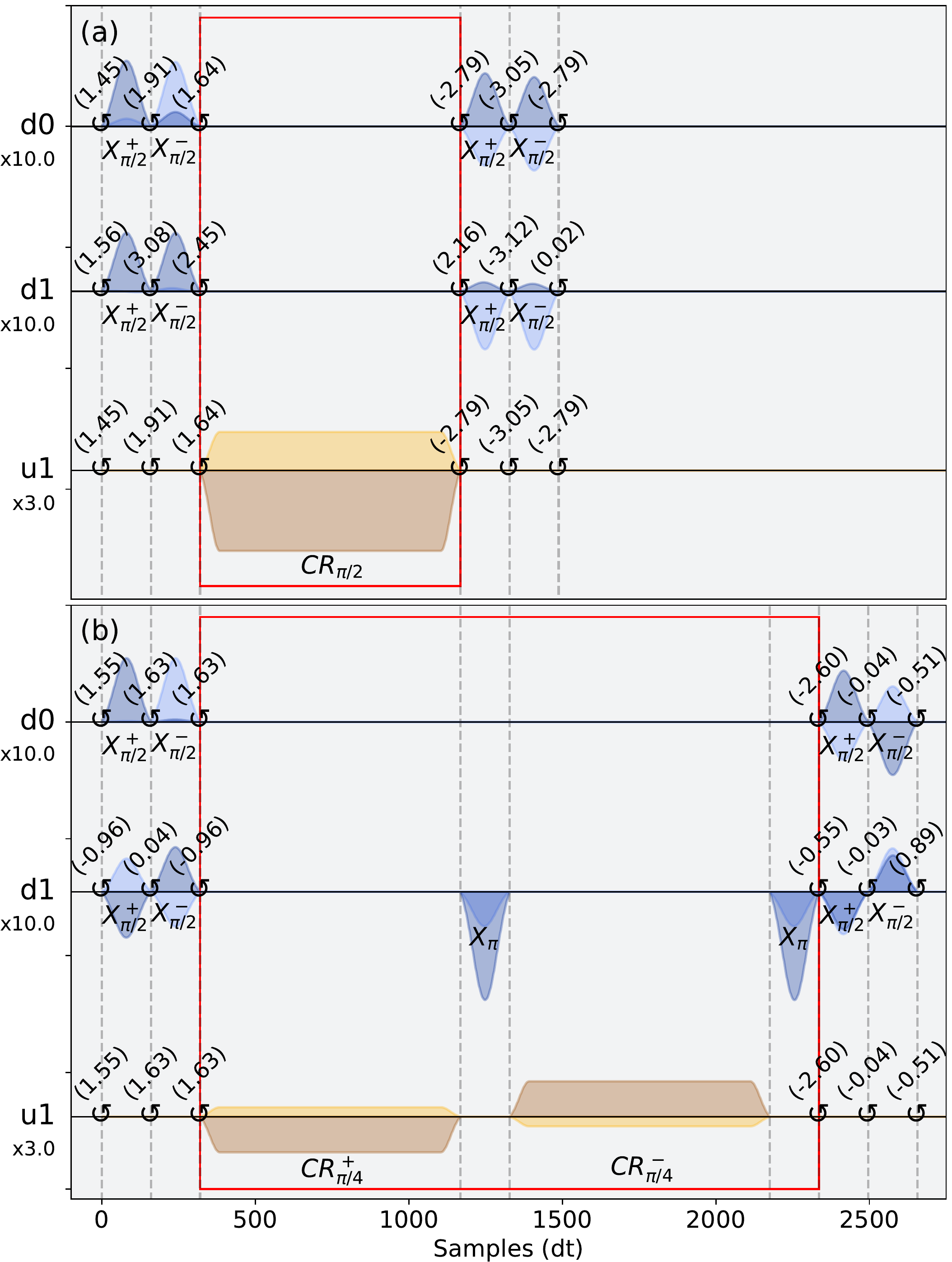}
  \caption{
  CNOT pulse \texttt{Schedule}s implemented by the CR gates with the local fidelity optimization. Local operations are realized by $X_{\pi/2}^\pm$ pulses with three virtual-Z gates before and after CR gates on the \texttt{DriveChannel}s \texttt{d0} and \texttt{d1}. The CR1 and CR2 gate are surrounded by red boxes. (a) CR1-based CNOT gate composed of a single CR pulse $CR_{\pi/2}$ on the \texttt{ControlChannel} \texttt{u1}. (b) CR2-based CNOT gate composed of two CR pulses $CR_{\pi/4}^\pm$ on \texttt{u1} with echo pulses $X_\pi^+$ applied on \texttt{d1}.
	Measurement and acquisition pulses are not shown.
  The numbers below the channel aliases show an amplitude scaling factor used for plotting.
  The 12 circular arrows topped by floating point numbers represent phase shifts in units of radians and each phase shift corresponds to an optimization parameter $\Theta_i$. Note that phase shifts on \texttt{u1} reflect those in \texttt{d0} to synchronize the frame of both channels; they are automatically inserted by the pulse scheduler.}
  \label{fig:cnot_schedules}
\end{figure}

The CR1 sequence is a single CR pulse on the \texttt{ControlChannel} \texttt{u1}, see Fig.~\ref{fig:cnot_schedules}(a).
The CR2 sequence consists of two CR pulses with opposite phases on \texttt{u1}, and two additional single-qubit pulses on the \texttt{DriveChannel} \texttt{d1}, one after each CR pulse, see Fig.~\ref{fig:cnot_schedules}(b).
This echo sequence refocuses unwanted terms in the interaction Hamiltonian \cite{Sheldon2016b}.
The following code exemplifies how to build the CR2 schedule in Qiskit Pulse.

\begin{minted}[fontsize=\footnotesize, frame=lines, framesep=2mm, autogobble]{python}
# Create a pulse schedule
sched = Schedule(name='cr2')

# Create pulse objects for the echoed CR gate
cr_p = GaussianSquare(cr_dur, amp, sigma, square_width)
cr_m = GaussianSquare(cr_dur, -amp, sigma, square_width)
x180 = Drag(pi_dur, pi_amp, pi_sigma, pi_beta)

# Assemble the schedule
sched += Play(cr_p, ControlChannel(1))
sched += Delay(t_cr, DriveChannel(1))
sched += Play(x180, DriveChannel(1))
sched += Play(cr_m, ControlChannel(1))
sched += Delay(t_cr, DriveChannel(1))
sched += Play(x180, DriveChannel(1))
\end{minted}

\subsection{Quantum Process Tomography of the CR Gate}\label{sec:qpt-cr-pulse}

To study the dynamics of the CR pulse, we perform standard QPT~\cite{Mohseni2008} of the CR1 and CR2 pulse sequences for a range of CR pulse amplitudes using the tomography module of Qiskit Ignis~\cite{qiskit-ignis}. Given a $d$-dimensional noisy quantum channel $\qmap$, QPT reconstructs the \emph{Choi}-matrix $\Lambda_{\qmap}$ which is the positive-semidefinite matrix defined by $\Lambda_{\qmap} \equiv \sum_{i=0}^{d-1}\ketbra{i}{i} \otimes\qmap(\ketbra{i}{i})$~\cite{Wood2015qic}.

\begin{figure}[htbp!]
  \includegraphics[width=0.95\linewidth]{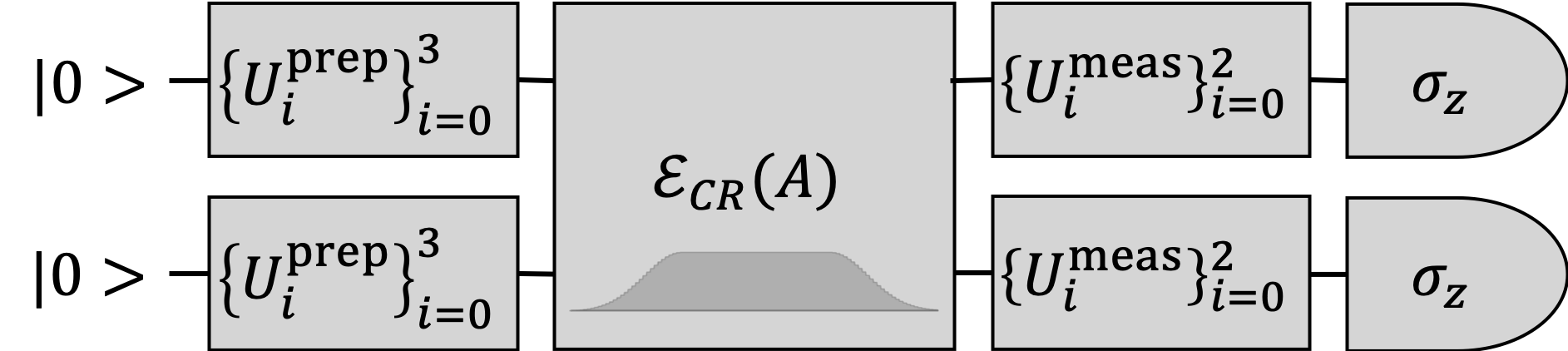}
  \caption{Process tomography circuits for the $\qmap_{\text{CR}}(A)$ pulse embedded as a user-defined custom gate.
  }
  \label{fig:process_tomography_diagram}
\end{figure}

The QPT circuits, shown in Fig.~\ref{fig:process_tomography_diagram}, use the single-qubit gates $\left\{U_i^{\textrm{prep}}\right\}_{i=0}^3$ and $\left\{U_i^{\textrm{meas}}\right\}_{i=0}^2$ to prepare the required input states and measurement bases, respectively.
We prepare each qubit in the states $\ket{0}, \ket{1}, \frac{1}{\sqrt{2}}(\ket{0} + \ket{1}), \frac{1}{\sqrt{2}}(\ket{0} +i \ket{1})$ and measure in the $X, Y$ and $Z$ bases.
Our amplitude-dependent CR pulse is inserted into the QPT circuits in Fig.~\ref{fig:process_tomography_diagram} as a custom gate that the Qiskit pulse scheduler maps to a pulse program, see appendix \ref{sec:qptprogramming}.
Each of the 144 two-qubit QPT pulse schedules is executed 2048 times to estimate the measurement outcome probabilities of each qubit.
The details of the readout process are presented in Sec.~\ref{sec:discriminators}.
We correct for measurement errors using the readout error mitigation techniques~\cite{Dewes2012} implemented in Qiskit Ignis.
Readout error mitigation for two qubits requires four additional schedules which were interleaved with the QPT schedules.
The reconstructed Choi-matrix $\qmap_{\text{CR}}(A)$ for the noisy gate was obtained from the convex-optimization QPT fitter in Qiskit Ignis for each value of the CR pulse amplitude $A$.
This fitter uses maximum likelihood estimation to find the completely-positive trace-preserving process that is most likely to fit the measured data after correction for readout errors.

\begin{figure}[htbp!]
  \includegraphics[width=0.95\linewidth]{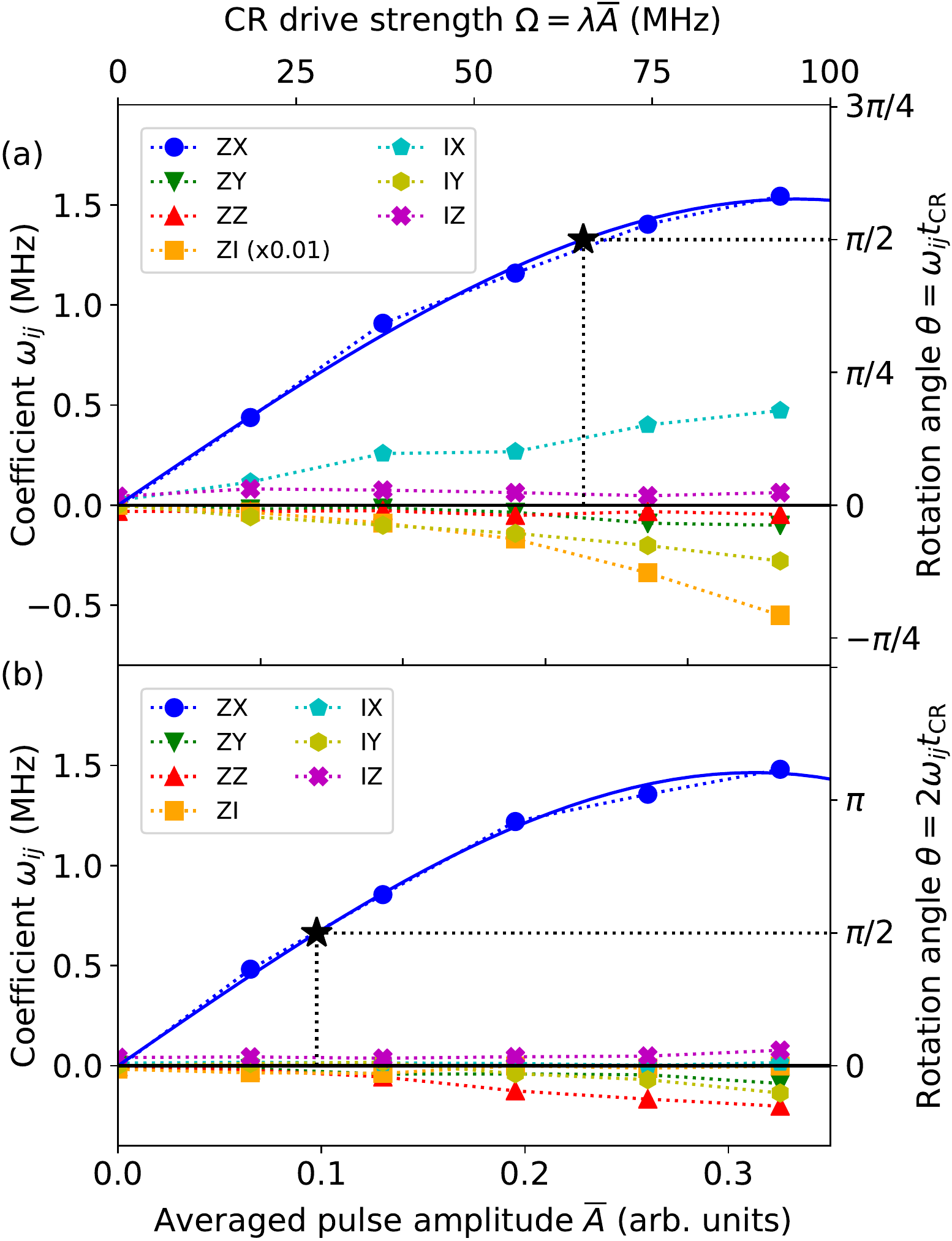}
  \caption{CR Hamiltonian coefficients reconstructed following the procedure described in Appendix \ref{sec:h_tomo}, plotted against the time-averaged CR pulse amplitude $\overline{A}$.
  The blue solid lines are the fits to the third order expansion of the CR Hamiltonian in Eq.~\eqref{eq:third_order}.
  The rotation angle is estimated by $\theta = n_{CR} \omega_{ij} t_\text{CR}$ where $n_{CR}$ is the number of CR pulses.
  Star-shaped markers indicate operating points of CNOT gates.
  (a) Measured coefficients of the CR1 sequence.
  The $ZI$ coefficient is several orders of magnitude larger than other Hamiltonian terms and is displayed with a 1/100 scale in (a) whereas in (b) it is approximately zeroed by the echo pulse in CR2 and is therefore not scaled.
  }
  \label{fig:cr_hamiltonian_qpt}
\end{figure}

We use the fitted Choi-matrices $\qmap_{\text{CR}}(A)$ to compute estimates of the coefficients of the effective CR Hamiltonian in Eq.~\eqref{eqn:hcr}.
The method is described in Appendix \ref{sec:h_tomo}.
We then fit these coefficients to a third order model to find the CR pulse amplitude that implements a $\theta_{ZX} = \pi/2$ rotation, see appendix \ref{sec:h_tomo}.
The estimated amplitudes, marked by the stars in Fig.~\ref{fig:cr_hamiltonian_qpt}, were $0.229 \pm 0019$ and $0.098 \pm 0005$ for CR1 and CR2, respectivly.

\subsection{Optimizing CNOT Fidelity with Local Operations}\label{sec:fid}\label{sec:cx-local-fidelity-optimization}

To estimate the highest fidelity of a maximally entangling gate that the CR gate can be transformed into, we optimize the average gate fidelity $F$ over all single-qubit pre and post-rotation angles $\Theta$ on both the control and target qubits, see appendix \ref{sec:local-fidelity-optimization}.
The optimized fidelities for the measured CR1 and CR2 process maps are $F_{\text{max}}=0.992$ and $F_{\text{max}}=0.994$.

We then use the Qiskit transpiler and pulse scheduler to optimize and build a CNOT gate from the calibrated CR gate and the device-calibrated single-qubit gates \footnote{The general single-qubit gate is built from a pair of $\pi/2$-pulses with three virtual Z-gates with rotation angles $\Theta_i$, $\Theta_j$, and $\Theta_k$.} which implement the optimal local rotation parameters $\Theta$ from Eq.~\eqref{eq:fid-opt}.
The optimized CNOT schedules are shown in Fig.~\ref{fig:cnot_schedules}.
The average gate fidelities of the calibrated CNOT gates are measured with randomized benchmarking (RB) \cite{magesanScalableRobustRandomized2011}.
The details of the pulse program applying the local rotations and setting up the RB measurements are given in Appendix \ref{sec:rbprogramming}.
The RB experiments estimate an average gate fidelity of $F=0.981$ and $F=0.979$ for the CR1 and CR2 gates, respectively.
These fidelities are comparable to the measured fidelity of $F=0.984$ of the standard CNOT gate provided by \texttt{ibmq\_almaden} which is implemented using a highly-tuned calibration process including an echo sequence, cancellation tone, and closed-loop amplitude calibration \cite{Sheldon2016b}.
It is worth noting that the CNOT gates demonstrated in this paper have no cancellation tone and all parameters are obtained with open-loop calibration.
In the same way, we can create custom basis gates which may enable hardware-efficient implementations of quantum algorigthms.

\section{Readout at the pulse level\label{sec:discriminators}}

Readout is the process through which the qubit state is projected onto $\ket{0}$ or $\ket{1}$ and a corresponding classical bit is obtained.
This process is modeled by a readout chain in which the observed signal undergoes a series of successive transformations. Qiskit Pulse supports returning the output data of each measurement layer to the programmer.
The lowest level accessible to the user, level-zero or \textit{raw} data, typically corresponds to a digitized time-series signal.
A kernel method applied to the signal data removes its time dependency and results in a complex value which encodes the qubit state (level-one \textit{kerneled} data).
Finally, the classified qubit state (level-two \textit{disciminated} data), is obtained by applying a discriminator to the kerneled data.
For a superconducting qubit processor, the time traces are complex vectors representing the digitized readout signals reflected or transmitted from the readout resonators \cite{Krantz2019}.
The kernel method, such as the boxcar integrator used within this paper, outputs points in the IQ plane which a discriminator may use to classify the qubit's state.

Qiskit users are now able to retrieve data from different levels in the readout-chain by specifying the readout data-level, i.e., zero, one, or two.
For example, users of IBM Quantum processors may request the kerneled data in the form of IQ points so that they may implement their own discriminator.
To be sure, Qiskit users that do not wish to implement their own kernels and discriminators can request level-two data therefore using the built-in readout scheme.
The readout methodology that we implemented reflects the typical data flow during readout in hardware and should allow users to test novel readout schemes \cite{Magesan2015} as well as accommodate different quantum computing architectures.

\begin{figure}[htbp!]
  \includegraphics[width=0.99\linewidth]{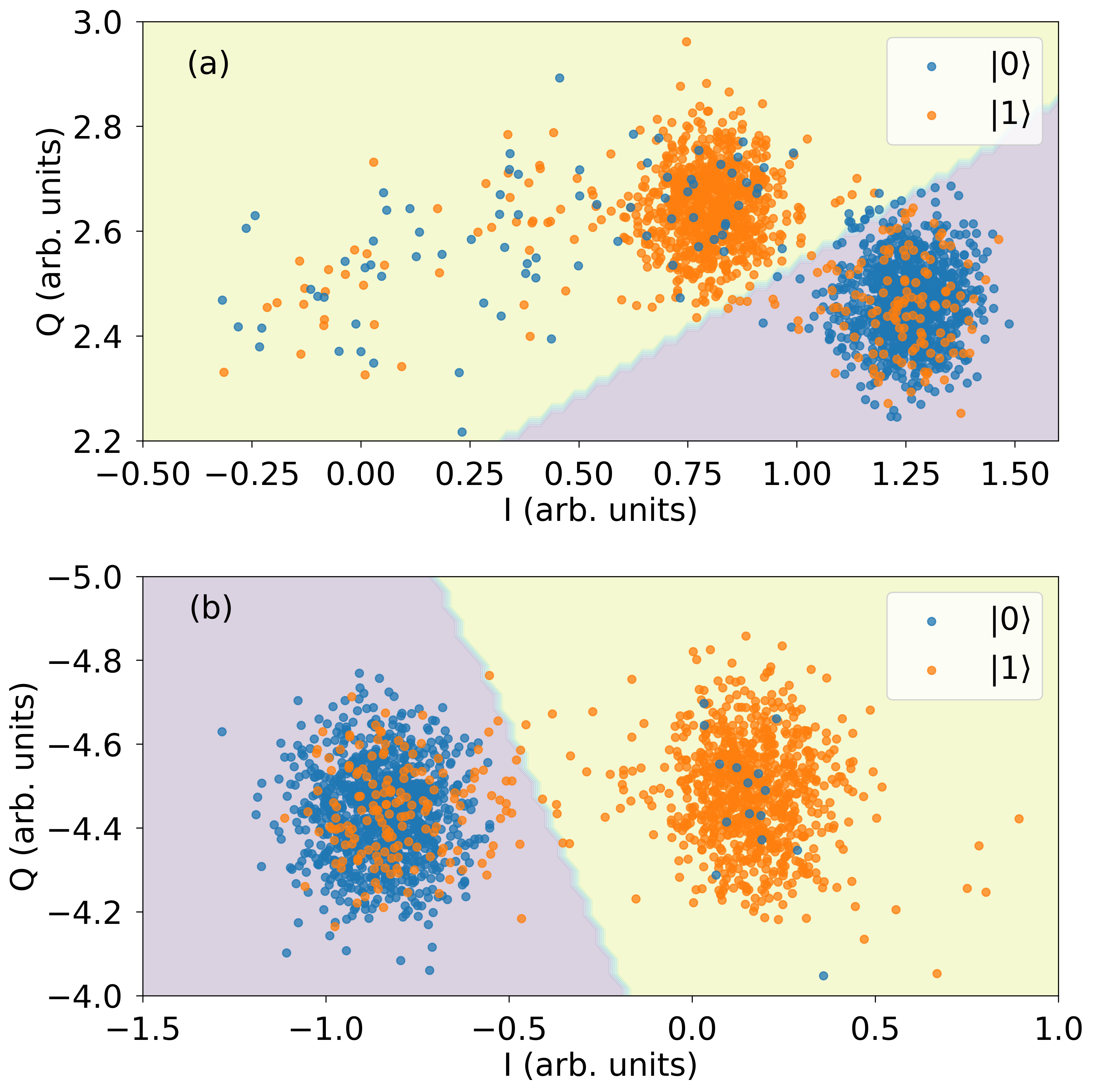}
  \caption{\label{fig:discriminator} IQ data and the decision boundary for the discriminators of qubits 16, shown in (a), and 17, shown in (b), that use only the IQ data of their respective qubit.}
\end{figure}

Aside from counting qubit states, discriminators may also be used to infer properties of the system and benchmark it.
We illustrate this by investigating spurious correlations in the qubit readout of the IBM Quantum system \texttt{ibmq\_singapore} selected based on its availability at the time of the experiment.
From \texttt{ibmq\_singapore} we randomly selected qubits 16 and 17 to study as they are a neighboring pair of qubits on the chip.
Kerneled data was measured for four calibration schedules, named \texttt{cal\_ij} with $i,j\in\{0,1\}$.
Here, to prepare the state $\ket{ij}$ a $\pi$-pulse is applied to qubit $i$ if $i=1$ and simultaneously to qubit $j$ if $j=1$.
The single-qubit pulses are followed by the measurement stimulus pulses and acquisition instructions for all qubits.
For each qubit we fit two discriminators based on linear discriminant analysis \cite{Magesan2015}.
For qubit $i$, one discriminator is fitted to a subset of the calibration data in which the other qubit is always in state $\ket{0}$, shown in Fig.~\ref{fig:discriminator}, while the other discriminator is fitted using all four calibration schedules.
We expect that the discriminator that uses all the calibration data will perform best if there is measurement cross-talk.
The fidelities of the fitted discriminators, shown in Tab.~\ref{tab:discrim}, suggest that there is no significant cross-talk between the qubits that we measured.
This is verified by $t$-tests on sixteen Pearsons' correlation coefficients $r_j(ES_{i,X}, GS_{i,Y})$ between $ES_{i,X}$ and $GS_{i,Y}$ which correspond to the $X,Y\in\{\rm{I},\rm{Q}\}$ data of qubit $i$ in state $j\in\{0,1\}$ when the other qubit is in the excited state ($ES$, i.e., $\ket{1}$) and ground state ($GS$, i.e., $\ket{0}$), respectively.
These correlation coefficients are sensitive to cross-talk between the two qubits.
With 1024 degrees of freedom, i.e., measurement shots, we do not observe any statistically significant correlation at the 95\% confidence level.
This implies that it is sufficient to fit discriminators using only a ground and excited reference schedule for each qubit and consequently only $2n$ calibration schedules are required for $n$ qubits when there is no cross-talk rather than the $2^n$ calibration schedules that would be required with all-to-all measurement crosstalk.

\begin{table}
 \centering
 \caption{\label{tab:discrim} The measurement assignment fidelity $F_a = 1 - (\Pr[0|1]+\Pr[1|0])/2$ of the four discriminators \cite{Magesan2015}. For example, the Single-Q16 discriminator was fitted with the calibration schedules \texttt{cal\_00} and \texttt{cal\_01} while the discriminator Both-Q16 was fitted with all four calibration schedules: \texttt{cal\_00}, \texttt{cal\_01}, \texttt{cal\_10} and \texttt{cal\_11}.
 The confidence intervals were obtained using Jeffreys interval at a 95\% confidence level.}
 \begin{tabular}{l l c c} \hline\hline
                             &        &\multicolumn{2}{c}{Qubit discriminated} \\
                             &        & Q16     & Q17 \\ \hline
  \multirow{2}{*}{Data used} & Single & $89.21^{+1.78}_{-2.00}\%$ & $91.06^{+1.42}_{-1.65}\%$ \\
                             & Both   & $89.48^{+1.26}_{-1.38}\%$ & $90.62^{+1.05}_{-1.16}\%$ \\ \hline\hline
 \end{tabular}
\end{table}

\section{Conclusions and Future Work\label{sec:conclusion}}
Rapid development in quantum computing has led to publicly available quantum computers with an increasing number of qubits, improved connectivity, and greater control.
Prior to this work, publicly available quantum programming frameworks for cloud-based quantum computers have been at the relatively high-level of the circuit model, or implementation-specific, thus limiting their application.
In the near-term, pulse-level control is desired to extract as much quantum volume as possible from the hardware by experimenting with novel control and characterization schemes \cite{bishopQuantumVolume2017, White2019, werninghaus2020}.

We have introduced Qiskit Pulse, an implementation of the virtual pulse-level programming model, OpenPulse \cite{Mckay2018}.
We have demonstrated that the Qiskit circuit scheduler can target pulse instructions and that physical superconducting qubit hardware can interpret these instructions to execute useful programs.
By embedding our pulse programming instruction set in Qiskit we have integrated gate-level quantum programs and classical pulse stimulus, exposing a new level of hardware control to Qiskit users.
The benefit that pulse control provides quantum programmers was demonstrated by calibrating a cross-resonance pulse on a cloud-based quantum computer and embedding it as a gate within the standard circuit programming model and characterizing this user-defined gate using quantum process tomography.

Giving users pulse-level access to current-day quantum computers will allow them to explore techniques such as error mitigation and dynamical decoupling schemes that cannot be investigated at the circuit level.
In the future we will explore embedding the pulse programming model as a coprocessor within a classical virtual instruction set architecture that supports classical arithmetic and control-flow \cite{Adve2003}.
We will also investigate extensions to the pulse programming model such as defining special purpose registers to track phase across multiple channels which would reduce the number of required \texttt{PulseChannel}s and enable simpler tracking of shared phase for composite gates.
We would then use these capabilities to explore the implementation of active error-correcting codes, and promising variational quantum-classical algorithms such as the variational quantum eigensolver \cite{Noh2019, Kandala2017, Ganzhorn2019}.

\section*{Acknowledgements}
We thank Blake Johnson for insightful discussions. We thank Andrew Wack and John Lemek for providing system access and deployments.

\bibliographystyle{unsrt}
\bibliography{bibliography}

\appendix

\onecolumngrid
\section{Programming with Qiskit Pulse}
\label{sec:pulseprogramming}

\subsection{Performing Quantum Process Tomography}
\label{sec:qptprogramming}
The code example below demonstrates how the quantum process tomography (QPT) schedules for the CR1 used in Sec.~\ref{sec:cr-pulse} are created with Qiskit Pulse.

\begin{minted}[frame=lines, linenos]{python}
from qiskit import transpile, schedule
from qiskit.circuit import QuantumRegister, QuantumCircuit, Gate
from qiskit.pulse import Schedule, Play, ControlChannel
from qiskit.pulse.pulse_lib import GaussianSquare
from qiskit.ignis.verification.tomography import process_tomography_circuits
from qiskit.test.mock import FakeAlmaden

# Unique name for the instruction.
gate_name = 'cr1'

# Pulse parameters determined during calibration.
duration = 848
sigma = 32
square_width = 64
amp = 0.2

# Call the backend and get basis_gates and inst_map.
backend = FakeAlmaden()
config = backend.configuration()
defaults = backend.defaults()
basis_gates = config.basis_gates
inst_map = defaults.instruction_schedule_map

# Create the CR1 schedule.
cr1_pulse = GaussianSquare(duration, amp, sigma, square_width)
sched = Schedule()
sched += Play(cr1_pulse, ControlChannel(0))

# Add the CR1 instruction to basis_gates and inst_map.
basis_gates += [gate_name]
inst_map.add(gate_name, [1, 0], sched)

# Create a quantum gate to reference the CR1 pulse schedule.
cr1_gate = Gate(gate_name, 2, [])

# Create the QPT circuits.
qregs = QuantumRegister(config.n_qubits)
circuit = QuantumCircuit(qregs)
circuit.append(cr1_gate, qargs=[qregs[1], qregs[0]])
qpt_circuits = process_tomography_circuits(circuit, [qregs[0], qregs[1]])

# Create the QPT pulse schedules.
qpt_circuits = transpile(qpt_circuits, backend, basis_gates)
qpt_schedules = schedule(qpt_circuits, backend, inst_map)
\end{minted}

The above example uses the mock backend \texttt{FakeAlmaden} for IBM Quantum system \texttt{ibmq\_almaden} which can be substituted for the real backend.

The pulse envelope of the CR1 \texttt{cr1\_pulse} is created with a flat-topped Gaussian pulse.
The pulse schedule \texttt{sched} of CR1 is then added to the \texttt{basis\_gates} and the circuit instruction to pulse schedule mapping (\texttt{inst\_map}) for qubits one and zero with the name \texttt{cr1}.
The \texttt{basis\_gates} defines a list of primitive circuit instructions available in the system and the \texttt{inst\_map} defines a lookup table of calibrated pulse schedules for each basis gate on each qubit.
The \texttt{circuit} object of the CR1 sequence is created with a custom gate \texttt{cr1\_gate}.
The QPT circuits are then assembled by calling the \texttt{process\_tomography\_circuits} function in Qiskit-Ignis.
This appends state preparation circuits and measurement circuits before and after the \texttt{circuit}.
The returned \texttt{qpt\_circuits} is a list of quantum circuits containing each possible combination of input states and measurement bases.
The \texttt{qpt\_circuits} are then mapped to the backend in question by Qiskit's transpiler, taking into account the extended set of \texttt{basis\_gates}.
Finally, we call the pulse scheduler with the custom \texttt{inst\_map} containing CR1 to create the QPT pulse schedules.
The QPT program for the CR2 sequence is created with the same procedure.

\subsection{Performing Randomized Benchmarking}
\label{sec:rbprogramming}
The code example below demonstrates how the standard randomized benchmarking (RB) schedules with the CR1-CNOT used in Sec.~\ref{sec:cx-local-fidelity-optimization} are created in the qiskit pulse module.

\begin{minted}[frame=lines, linenos]{python}
from qiskit import transpile, schedule
from qiskit.circuit import QuantumCircuit, Gate
from qiskit.pulse import Schedule, Play, ControlChannel
from qiskit.pulse.pulse_lib import GaussianSquare
from qiskit.ignis.verification.randomized_benchmarking import randomized_benchmarking_seq
from qiskit.test.mock import FakeAlmaden

# Unique name for instruction.
gate_name = 'cr1'

# Randomized benchmark setup.
n_seed = 5
n_clifford = [1, 21, 51, 101, 151]

# Pulse parameters determined during calibration.
duration = 848
sigma = 32
square_width = 64
amp = 0.2

# Local rotation angles for CNOT(1,0) determined during optimization.
local_rotations10 = [[1.45, 1.91, 1.64],
                     [1.56, 3.08, 2.45],
                     [-2.79, -3.05, -2.79],
                     [2.16, -3.12, 0.02]]

# Local rotation angles for CNOT(0,1) determined during optimization.
local_rotations01 = [[-1.68, 3.04, 1.66],
                     [1.57, 2.28, -0.06],
                     [1.48, -0.46, 3.14],
                     [1.60, -3.14, 0.98]]

# Call the backend and get the basis_gates and inst_map.
backend = FakeAlmaden()
config = backend.configuration()
defaults = backend.defaults()
basis_gates = config.basis_gates
inst_map = defaults.instruction_schedule_map

# Create the CR1 schedule.
cr1_pulse = gaussian_square(duration, amp, sigma, square_width)
sched = Schedule()
sched += Play(cr1_pulse, ControlChannel(0))

# Add the CR1 instruction to basis_gates and inst_map.
basis_gates += [gate_name]
inst_map.add(gate_name, [1, 0], sched)

# Create a quantum gate to reference the CR1 pulse schedule.
cr1_gate = Gate(gate_name, 2, [])

# Build a CNOT(1,0) schedule based on CR1 schedule.
qregs = QuantumRegister(config.n_qubits)
cnot10 = QuantumCircuit(qregs)
cnot10.u3(*local_rotations10[0], qregs[0])
cnot10.u3(*local_rotations10[1], qregs[1])
cnot10.append(cr1_gate, qargs=[qregs[1], qregs[0]])
cnot10.u3(*local_rotations10[2], qregs[0])
cnot10.u3(*local_rotations10[3], qregs[1])

cnot10 = transpile(cnot10, backend, basis_gates)
cnot_sched10 = schedule(cnot10, backend, inst_map)

# Build a CNOT(0,1) schedule based on CR1 schedule.
qregs = QuantumRegister(config.n_qubits)
cnot01 = QuantumCircuit(qregs)
cnot01.u3(*local_rotations01[0], qregs[0])
cnot01.u3(*local_rotations01[1], qregs[1])
cnot01.append(cr1_gate, qargs=[qregs[1], qregs[0]])
cnot01.u3(*local_rotations01[2], qregs[0])
cnot01.u3(*local_rotations01[3], qregs[1])

cnot01 = transpile(cnot01, backend, basis_gates)
cnot_sched01 = schedule(cnot01, backend, inst_map)

# Overwrite the default CNOT schedule in the inst_map.
inst_map.add('cx', [1, 0], cnot_sched10)
inst_map.add('cx', [0, 1], cnot_sched01)

# Create randomized benchmarking circuits with 5 seeds.
rb_circuits_seeds, _ = randomized_benchmarking_seq(n_seed, n_clifford, [[0, 1]])

# Schedule the randomized benchmarking experiment into pulse schedules.
rb_schedules_seeds = []
for rb_circuits_seed in rb_circuits_seeds:
    rb_circuits_seed = transpile(rb_circuits_seed, backend, basis_gates)
    rb_schedules_seed = schedule(rb_circuits_seed, backend, inst_map)
    rb_schedules_seeds.append(rb_schedules_seed)
\end{minted}

As shown in Sec.~\ref{sec:qptprogramming}, the pulse schedules are programmed with the aid of the \texttt{QuantumCircuit} class to apply device calibrated single-qubit gates around the CR1 pulse sequence abstracted by \texttt{cr1\_gate}.
The local rotation parameters can be obtained by the optimization routine shown in Sec.~\ref{sec:local-fidelity-optimization}.
It should be noted that in a two qubit standard RB sequence the CNOT gate can assign both qubit 0 and 1 as a control qubit.
Beacuse the CNOT gate is not identical under the exchange of the control and the target qubits , we need to prepare pulse schedules for both qubit arrangements.
Then, the default CNOT instruction in the \texttt{inst\_map} is overwitten by the pulse schedules based on the calibrated CR1 sequence.
Finally, the RB circuits are generated by a call to the \texttt{randomized\_benchmarking\_seq} function in Qiskit-Ignis.
The returned \texttt{rb\_circuits\_seeds} is a list of RB circuits for each random seed.
These RB circuits are then independently transpiled and scheduled to create RB pulse programs.
The RB programs for the CR2 sequence are created with the same procedure.

\section{Cross Resonance Phase Calibration\label{sec:phase_cal}}

In the twenty-qubit IBM Quantum system \texttt{ibmq\_almaden}, microwave pulses programmed with Qiskit Pulse are generated by waveform generators at room temperature and travel through coaxial cables to the qubits~\cite{Krinner2019}.
The transfer function between the room temperature electronics and the qubits can cause a phase offset $\phi_0$ in Eq.~\eqref{egn:pwcc_pulse} resulting in an error in the rotation axis of the target qubit.
The Hamiltonian may therefore have an unwanted $ZY$ interaction term which we eliminate by adjusting the phase of CR pulse $\phi$.
We perform this calibration with the CR2 schedule since its time-independent Hamiltonian, which we approximate by
\begin{equation}\label{eqn:approx_cr_with_zy}
\overline{H}_{\text{CR}} \simeq \Omega({A, \phi}) \left( \cos \phi_0 ZX + \sin \phi_0 ZY \right) + \varepsilon,
\end{equation}
has less terms than the CR1 Hamiltonian due to the echo.
Here, $\Omega$ is the strength of the CR drive as a function of its amplitude $A$ and its phase $\phi$ while $\varepsilon$ represents the small interaction terms which are not fully refocused by the echo sequence.

First, we initialize the qubit in the $\ket{00}$ state.
We sweep the amplitude $A$ and measure the target qubit in the Pauli-$Z$ basis to find the pulse amplitude $A_{\rm opt} = 0.108$ which creates an equal superposition of $\ket{0}$ and $\ket{1}$.
If the offset $\phi_0$ is zero this transformation is a $\pi/2$-rotation around the $X$-axis so that the target qubit, measured in the $Y$-basis, yields $\Tr(\opy\rho) = \pm1$ with the sign depending on the state of the control qubit.
We thus measure the readout signal at $A_{\rm opt}$ in the $Y$-basis for both initial states of the control qubit $\ket{10}$ and $\ket{00}$, while sweeping the phase $\phi$.
The calibrated phase that maximizes $\vert\Tr(\opy\rho)\vert$ is $\phi_{\rm opt} = -0.166~\rm{rad.}$

\section{Effective Hamiltonian Estimation and Amplitude Calibration\label{sec:h_tomo}}

We use the fitted Choi-matrices $\qmap_{\text{CR}}(A)$ to compute estimates of the coefficients of the effective CR Hamiltonian in Eq.~\eqref{eqn:hcr}.
Since a real CR pulse will have noise the resulting process is not unitary. Noisy quantum evolution for a time-independent Hamiltonian in the presence of Markovian noise may be described by the \emph{Lindblad equation} $\frac{d}{dt}\rho(t) = \mathcal{G}(\rho)$ with the Lindblad generator
\begin{align}
\mathcal{G}(\rho) &= \mathcal{L}_H(\rho) + \mathcal{D}(\rho) \\
\mathcal{L}_H(\rho) &= -i [H, \rho]  \\
\mathcal{D}(\rho) &= \sum_j \gamma_j \left(A_j \rho A_j^\dagger - \frac{1}{2}\{A_j^\dagger A_j, \rho\}\right),
\end{align}
where the operator $\mathcal{L}$ is the generator of the unitary evolution, and $\mathcal{D}$ is the generator of the non-unitary dissipative evolution. As with unitary evolution, the Lindblad equation can be solved as a matrix differential equation obtaining  $|\rho(t)\rangle\!\rangle = S_{\mathcal{E}}|\rho(0)\rangle\!\rangle$, where $|A\rangle\!\rangle$ denotes a \emph{column-vectorized} matrix $A$, and $S_{\mathcal{E}}= \exp(t S_{\mathcal{G}})$ is the \emph{superoperator} representation of quantum process $\mathcal{E}$~\cite{Wood2015qic}.

For a Hamiltonian $H$ we note that the operators $B_{ij} = \frac{1}{2} P_i\otimes P_j$, with $P_i$ are single-qubit Pauli operators, define an orthonormal basis for two-qubit operators --- i.e. $\Tr[B_{ij} B_{kl}^\dagger] = \delta_{ik}\delta_{jl}$. Hence for a Hamiltonian given by $H = \sum_{ij} \omega_{ij} B_{ij}$, we can extract the coefficients via $\omega_{ij}=Tr[B_{ij}^\dagger H]$. The superoperator for the Hamiltonian component of $\mathcal{G}$ is given by
\begin{equation}
S_{\mathcal{L}_H} = -i(\mathbb{I}\otimes H - H^T \otimes\mathbb{I}).
\end{equation}

We can use the fact that the superoperators of the Hamiltonian basis term $S_{\mathcal{L}_{B_{ij}}}$ are also an orthogonal (but not-normalized) basis for $S_{\mathcal{L}_H}$, and importantly, are orthogonal to the dissipative part of the generator ($\Tr[S_{\mathcal{L}_{B_{ij}}}^\dagger S_{\mathcal{D}}]=0$) when the dissipator only involves Pauli and $T_1$ and $T_2$ relaxation terms. This allows us to extract the coefficients from the Lindblad superoperator generator as
\begin{align}\label{eqn:cr_ham_reconstruction}
\omega_{ij} = \frac{\Tr \left[S_{\mathcal{L}_{B_{ij}}}^\dagger  S_{\mathcal{G}} \right]}{\| S_{\mathcal{L}_{B_{ij}}} \|}.
\end{align}

To compute the superoperator generator $S_{\mathcal{G}}$, we first obtain the Choi-matrix estimate for a channel $\mathcal{E}$ from quantum process tomography and then convert it to the superoperator representation $S_{\mathcal{E}}$. For additional details on the superoperators and converting between superoperators and the Choi-matrix representation obtained from tomography see \cite{Wood2015qic}. Next, we take the matrix logarithm to obtain the generator $S_{\mathcal{G}} = t^{-1}\log(S_{\mathcal{E}})$ from which we estimate $\omega_{ij}$ for our two-qubit system using Eq.~\eqref{eqn:cr_ham_reconstruction}.
The process fidelities of the estimated CR Hamiltonian using this technique and the experimentally obtained Choi-matrix are 99.4 \% and 98.6 \% on average for CR1 and CR2 experiments, respectively.


We find that as predicted only the terms $ZX$, $ZY$, $ZZ$, $ZI$, $IX$, $IY$, and $IZ$, shown in Fig.~\ref{fig:cr_hamiltonian_qpt}, are significant for CR1 and CR2 while all other remaining Pauli terms are negligible.
In both CR sequences the $ZY$ term is suppressed by the calibrated CR phase $\phi_{\rm opt}$ and a monotonic increase of the desired $ZX$ term is observed as the pulse amplitude $A$ increases.
The CR1 pulse without echoing has large contributions from the $IX$, $IY$ and $ZI$ terms, see Fig.~\ref{fig:cr_hamiltonian_qpt}(a).
Such unwanted interactions, except for $ZZ$, are removed by the echo sequence in CR2, compare Fig.~\ref{fig:cr_hamiltonian_qpt}(a) and (b).
The effect of these unwanted interactions can be reduced by applying single-qubit gates before and after the CR pulse to correct for local coherent errors as discussed in Sec.~\ref{sec:fid}.

We now find the CR pulse amplitude that creates a maximum entangling gate, i.e. $\theta_{ZX} = \pi/2$ in Eq.~\eqref{eqn:ideal_cr}.
Due to the Gaussian edges of our CR pulses we relate the drive strength $\Omega$ to the time-averaged pulse amplitude $\overline{A}$ through a linear response $\Omega=\lambda \overline{A}$.
The measured $ZX$ interaction strengths are fit by the third order expansion of the CR Hamiltonian~\cite{Magesan2018}
\begin{align} \label{eq:third_order}
\frac{\omega_{ZX}(\overline{A})}{2} =& -\frac{J\lambda\overline{A}}{\Delta}\left(\frac{\delta_1}{\delta_1+\Delta}\right) \\ \notag
& \hspace{-5mm} + \frac{J(\lambda\overline{A})^3\delta_1^2(3\delta_1^3+11\delta_1^2\Delta+15\delta_1\Delta^2+9\Delta^3)}{4\Delta^3(\delta_1+\Delta)^3(\delta_1+2\Delta)(3\delta_1+2\Delta)}
\end{align}
where $J$ is the coupling strength, $\delta_1$ is the anharmonicity of the control qubit, and $\Delta$ is the frequency difference between the control and target qubits. In this model, we have a pair of fit parameters $J$ and $\lambda$.
The coupling strength obtained from the fit was $J = 1.87 \pm 0.046~\rm{MHz}$ and $1.79 \pm 0.033~\rm{MHz}$ for the CR1 and CR2 data, respectively.
The $\lambda$ coefficient was $-271.2 \pm 12.2~\rm{MHz}$ and $-288.9 \pm 10.2~\rm{MHz}$ for the CR1 and CR2 data, respectively.
These fit parameters are independent of the pulse sequences and both results almost agree within the error range.
The small mismatch between the fit values may be caused by imperfections in the Hamiltonian reconstructed from the tomography data.
These fit curves yield the controlled rotation angle as a function of the CR pulse amplitude $\theta_{ZX}(\overline{A}) = n_{CR} \omega_{ZX}(\overline{A}) t_\text{CR}$ where $n_{CR}=1$ for CR1 and $n_{CR}=2$ for CR2. Finally, we can find the pulse amplitudes $\overline{A}$ for $\theta_{ZX} = \pi/2$. The estimated amplitudes were respectivly $0.229 \pm 0019$ and $0.098 \pm 0005$ for CR1 and CR2. These are marked by the stars in Fig.~\ref{fig:cr_hamiltonian_qpt}.
Due to the nonlinearity between the $ZX$ term and the average pulse amplitude $\overline{A}$, see Eq.~(\ref{eq:third_order}), the estimated drive amplitude of CR1 is slightly larger than double the drive amplitude of CR2.



\section{Optimizing CNOT Fidelity with Local Operations}\label{sec:local-fidelity-optimization}

To estimate the highest fidelity of a maximally entangling gate that the CR gate can be transformed into, we optimize the average gate fidelity $F$ over all single-qubit pre and post-rotations on both the control and target qubits:
\begin{equation}
F_{{\text{max}}} = \max_{\Theta} F[\qmap_{CR}(\overline{A}_{\pi/2}), U_{\text{ent}}(\Theta)] \label{eq:fid-opt}
\end{equation}
where the optimization is over 12 real parameters for the four parameterized $U_3(\Theta_i, \Theta_j, \Theta_k)$ rotations in $SU(2)$:
\begin{align} \notag
U_{\text{ent}}(\Theta) & = U_{\text{pre}}^\dagger(\Theta) U_{\text{ent}} U_{\text{post}}^\dagger(\Theta),\\ \notag
U_{\text{pre}}(\Theta) & = U_3(\Theta_0, \Theta_1, \Theta_2) \otimes U_3(\Theta_3, \Theta_4, \Theta_5),\\ \notag
U_{\text{post}}(\Theta) & = U_3(\Theta_6, \Theta_7, \Theta_8) \otimes U_3(\Theta_9, \Theta_{10}, \Theta_{11}).
\end{align}
The ideal unitary matrix for the cross-resonance perfect-entangler is $U_{\text{ent}}=ZX(\pi/2)$ and for the CNOT gate is $U_{\text{ent}}=CX$.
This optimization aims to remove the effect of locally correctable coherent errors.
It will, however, underestimate the error of the transformed CR gate as it neglects errors in the single-qubit gates.
The optimized fidelities for the measured CR1 and CR2 process maps are $F_{\text{max}}=0.994$ and $F_{\text{max}}=0.998$.

\end{document}